\documentclass[aps,superscriptaddress,twocolumn,showpacs,preprintnumbers,amsmath,amssymb]{revtex4}
\usepackage[cp1251]{inputenc}
\usepackage{amssymb,amsmath}
\usepackage[mathscr]{eucal}
\usepackage[pdftex]{graphicx}
\usepackage{longtable}
\usepackage[unicode, colorlinks=false,linkcolor=blue, pdfborder={0 0 0}]{hyperref} 
\begin{document}
\title{Dual Symmetric Solution of Maxwell Equations and Correct Quantization of  Electromagnetic Field}
\author{A.Dovlatova}
\affiliation{M.V.Lomonosov Moscow State University, Moscow, 119899, alex@ph-elec.phys.msu.su}
\author{D.Yearchuck}
\affiliation{Minsk State Higher College, Uborevich Str.77, Minsk, 220096, RB; yearchuck@gmail.com}
\author{Y.Yerchak}
\affiliation{Belarusian State University, Nezavisimosti Ave.4, Minsk, 220030, RB; Jarchak@gmail.com}
\author{A.Alexandrov}
\affiliation{M.V.Lomonosov Moscow State University, Moscow, 119899, alex@ph-elec.phys.msu.su}
\date{\today}
\begin{abstract}It has been found, that free electromagnetic (EM) field in restricted volume (typical experimental case) consists of two independent and equally possible components with different parity under spatial inversion transformations. Either of  the two  components indicated represents the system  of also two independent and equally possible fields, which are even and uneven under time reversal transformations. The rules for local quantization of EM-field in Minkowski space are obtained.
 
\end{abstract}
\maketitle
\section{Introduction and Background}
In 1873 Y  "A Treatise on Electricity and Magnetism" by Maxwell  \cite{Maxwell} 
was published, in which the discovery of the system of electrodynamics equations was reported. The equations  are in fact the symmetry expressions for experimental laws, established by Faraday, and, consequently, they are mathematical mapping of experimentally founded   symmetry of EM-field. It means in its turn that if some new  experimental data will indicate, that symmetry of EM-field is higher, then Maxwell equations have to be generalized. That is the reason why the symmetry study of  Maxwell equations is the subject of many research in field theory up to now. Heaviside \cite{Heaviside} in twenty years after Maxwell discovery was the first, who payed attention to the symmetry between electrical and magnnetic quantities in Maxwell equations. Mathematical formulation of given symmetry, consisting in  invariance of Maxwell equations for free EM-field under the duality transformations
\begin{equation} 
\label{eq1d}
\vec {E} \rightarrow \pm\vec {H}, \vec {H} \rightarrow \mp\vec {E},
\end{equation}
gave Larmor \cite{Larmor}.
 Duality transformations (\ref{eq1d}) are private case of the more general dual transformations, established by Rainich \cite{Rainich}. Dual transformations produce oneparametric abelian  group $U_1$ of chiral transformations and they are 
\begin{equation} 
\label{eq2d}
\begin{split}
\raisetag{40pt}                                                    
\vec {E} \rightarrow \vec {E} cos\theta + \vec {H} sin\theta\\
\vec {H} \rightarrow \vec {H} cos\theta - \vec {E} sin\theta.
\end{split}
\end{equation}
Given symmetry indicates, that both constituents $\vec {E}$ and $\vec {H}$ of EM-field are possessing equal rights, in particular they both have to consist of component with different parity. Subsequent extension of dual symmetry for the EM-field with sources leads to requirement of two type of charges. Examples of the dual symmetry display are for instance the equality of magnetic and electric energy values in LC-tank or in free electromagnetic wave. Recently concrete experimental results have been obtained concerning dual symmetry of EM-field in the matter.
 Two new physical phenomena - ferroelectric \cite{Yearchuck_Yerchak} and antiferroelectric \cite{Yearchuck_PL} spin wave resonances have been  observed.    They were predicted on the base of the model \cite{Yearchuck_Doklady} for the chain of electrical "spin" moments, that is intrinsic electrical moments of (quasi)particles. Especially interesting, that in \cite{Yearchuck_PL} was experimentally proved, that really purely imaginary electrical "spin" moment, in full correspondence with Dirac prediction \cite{Dirac}, is responsible for the phenomenon observed. Earlier on the same samples has been registered ferromagnetic spin wave resonance, \cite{Ertchak_J_Physics_Condensed_Matter}.

The values of splitting parameters $\mathfrak{A}^E$ and $\mathfrak{A}^H$ in ferroelectric and ferromagnetic spin wave resonance spectra allowed to find 
 the ratio $J_{E }/J_{H}$ of exchange 
constants    in the range of $(1.2 - 1.6)10^{4}$. Given result seems to be direct proof,  that the charge, that is function, which is invariant under gauge transformations is two component function.  The ratio of imagine $e_{H} \equiv g$ to real $e_{E}\equiv e $  components of complex charge is $\frac{g}{e} \sim \sqrt{J_{E }/J_{H}} \approx (1.1 - 1.3)10^{2}$. At the same time in classical and in quantum theory dual symmetry of Maxwell eqations does not take into consideration. Moreover the known solutions of Maxwell eqations do not reveal given symmetry even for free EM-field, see for instance \cite{Scully}, although it is understandable, that the general solutions have to posseess by the same symmetry. 

The aim of given work is to find the cause of symmetry difference of Maxwell eqations  and their solutions and to propose correct field functions for classical and quantized EM-field. 
\section{Dual Symmetry Quantization of EM-field in volume rectangular cavity}  
Suppose EM-field in volume rectangular cavity. Suppose also, that the field polarization is linear in z-direction. Then the vector of electrical component can be represented in the form 
\begin{equation}
E_x(z,t) = \sum_{\alpha=1}^{\infty}A_{\alpha}q_{\alpha}(t)\sin(k_{\alpha}z),
\end{equation}
where $q_{\alpha}(t)$ is amplitude of $\alpha$-th normal mode of the cavity, $\alpha \in N$,
\begin{equation}
k_{\alpha} = \frac{\alpha\pi}{L}, A_{\alpha}=\sqrt{\frac{2 {\omega^2}_\alpha m_\alpha}{V\epsilon_0}}, \omega_{\alpha} = \frac{\alpha \pi c}{L},
\end{equation}
 $L$ is cavity length along z-axis, $V$ is cavity volume, $m_{\alpha}$ is parameter, which is introduced to obtain the analogy with mechanical harmonic oscillator. Using the equation
\begin{equation}
\epsilon_0\partial_t \vec{E}(z,t) = \left[ \nabla\times\vec{H}(z,t)\right]
\end{equation}
 we obtain the expression for magnetic field    
\begin{equation}
   {H}_y(z,t) =  \sum_{\alpha=1}^{\infty}\epsilon_0\frac{A_{\alpha}}{k_{\alpha}}\frac{dq_{\alpha}}{dt}\cos(k_{\alpha}z) + H_{y0}(t),
\end{equation}
where $H_{y0} = \sum_{\alpha=1}^{\infty} f_{\alpha}(t)$, $\{f_{\alpha}(t)\}$, $\alpha \in N$, is the set of arbitrary  functions of the time.
The partial solution is usually used, in which the function $H_{y0}(t)$ is identically zero.
The field Hamiltonian $\mathcal{H}^{[1]}(t)$, corresponding given partial solution, is
\begin{equation}
\begin{split}
&\mathcal{H}^{[1]}(t) = \frac{1}{2}\iiint\limits_{(V)}\left[\epsilon_0E_x^2(z,t)+\mu_0H_y^2(z,t)\right]dxdydz\\
&= \frac{1}{2}\sum_{\alpha=1}^{\infty}\left[m_{\alpha}\omega_{\alpha}^2q_{\alpha}^2(t) + \frac{p_{\alpha}^2(t)}{m_{\alpha}} \right],
\end{split}
\end{equation}
where
\begin{equation}
p_{\alpha} = m_{\alpha} \frac{dq_{\alpha}(t)}{dt}.
\end{equation}
Then, using the equation
\begin{equation}
\left[ \nabla\times\vec{E}\right] = -\frac{\partial \vec{B}}{\partial t} = -\mu_0 \frac{\partial \vec{H}}{\partial t}
\end{equation}
it is easily to find the field functions $\{q_{\alpha}(t)\}$. They will satisfy to  differential equation
\begin{equation}
\frac{d^2q_{\alpha}(t)}{dt^2}+\frac{k_{\alpha}^2}{\mu_0\epsilon_0}q_{\alpha}(t)=0.
\end{equation}
Consequently, taking into account $\mu_0\epsilon_0 = 1/c^2$, we have
\begin{equation}
q_{\alpha}(t) = C_1e^{i\omega_{\alpha}t}+C_2e^{-i\omega_{\alpha}t}
\end{equation}
 
Thus, real free Maxwell field equations result in well known in the theory of differential equations  
situation - the solutions are complex-valued functions. It means, that generally the field function for free Maxwell field produce complex space.

From general expression for the  field $\vec{H}(\vec{r},t)$ 
\begin{equation}
\vec{H}(\vec{r},t) =  \left[\sum_{\alpha=1}^{\infty}A_{\alpha}\frac{\epsilon_0}{k_{\alpha}}\frac{dq_{\alpha}(t)}{dt}\cos(k_{\alpha}z) + f_{\alpha}(t)\right]\vec{e}_y
\end{equation}
it  is easily to obtain differential equation  for $f_{\alpha}(t)$
\begin{equation}
\begin{split}
&\frac{d f_{\alpha}(t)}{dt} + A_{\alpha}\frac{\epsilon_0}{k_{\alpha}}\frac{\partial^2q_{\alpha}(t)}{\partial t^2}\cos(k_{\alpha}z) \\
&+ \frac {1}{\mu_0} A_{\alpha}k_{\alpha}q_{\alpha}(t)cos(k_{\alpha}z) = 0.
\end{split}
\end{equation}
Its solution
in general case is
\begin{equation}
f_{\alpha}(t) = \int\limits_{0}^{t} A_{\alpha} \cos(k_{\alpha}z)\left[- q_{\alpha}(\tau)\frac{k_{\alpha}}{\mu_0}-\frac{d^2q_{\alpha}(\tau)}{d\tau^2}\frac{\epsilon_0}{k_{\alpha}}\right]
d\tau 
\end{equation}
Then we have another solution of Maxwell equations 
\begin{equation}
\vec{H}^{[2]}(\vec{r},t) = \frac{(-1)}{\mu_0}\left\{\sum_{\alpha=1}^{\infty}k_{\alpha}A_{\alpha} \cos(k_{\alpha}z) q_{\alpha}'(t)\right\}\vec{e}_y,
\end{equation}
\begin{equation}
\vec{E}^{[2]}(\vec{r},t) = \left\{\sum_{\alpha=1}^{\infty}A_{\alpha}\frac{dq_{\alpha}'(t)}{dt}\sin(k_{\alpha}z)\right\}\vec{e}_x,
\end{equation}
where
$q_{\alpha}'(t) = \int\limits_{0}^{t} q_{\alpha}(\tau) d\tau$.
The Hamiltonian $\mathcal{H}^{[2]}(t)$ is
\begin{equation}
\mathcal{H}^{[2]}(t) = \frac{1}{2}\sum_{\alpha=1}^{\infty}\left[m_{\alpha} \omega_{\alpha}^4 q{'}_{\alpha}^2(t) + {m_{\alpha}\omega_{\alpha}^2 (\frac{dq_{\alpha}'(t)}{dt})^2} \right].
\end{equation}
Let us introduce new variables
\begin{equation}
\begin{split}
&q{''}_{\alpha}(t) = \omega_{\alpha}q{'}_{\alpha}(t) \\
&p{''}_{\alpha}(t) = m_{\alpha}\omega_{\alpha}\frac{dq_{\alpha}'(t)}{dt}
\end{split}
\end{equation}
Then
\begin{equation}
\mathcal{H}^{[2]}(t) = 
\frac{1}{2}\sum_{\alpha=1}^{\infty}\left[m_{\alpha}\omega_{\alpha}^2q{''}_{\alpha}^2(t) + \frac{p{''}_{\alpha}^2(t)}{m_{\alpha}} \right].
\end{equation}
We use further the standard procedure of field quantization. So for the first partial solution we have
\begin{equation}
\begin{split}
&\left[\hat {p}_{\alpha}(t) , \hat {q}_{\beta}(t)\right] = i\hbar\delta_{{\alpha}\beta}\\
&\left[\hat {q}_{\alpha}(t) , \hat {q}_{\beta}(t)\right] = \left[\hat {p}_{\alpha}(t) , \hat {p}_{\beta}(t)\right] = 0,
\end{split}
\end{equation}
where
$\alpha, \beta \in N$.
Introducing the operators $\hat{a}_{\alpha}(t)$ and $ \hat{a}^{+}_{\alpha}(t)$
\begin{equation}
\begin{split}
&\hat{a}_{\alpha}(t) = \frac{1}{ \sqrt{ 2 \hbar  m_{\alpha} \omega_{\alpha}}} \left[ m_{\alpha} \omega_{\alpha}\hat {q}_{\alpha}(t) + i \hat {p}_{\alpha}(t)\right]\\
&\hat{a}^{+}_{\alpha}(t) = \frac{1}{ \sqrt{ 2 \hbar  m_{\alpha} \omega_{\alpha}}} \left[ m_{\alpha} \omega_{\alpha}\hat {q}_{\alpha}(t) - i \hat {p}_{\alpha}(t)\right],
\end{split}
\end{equation}
we have for the operators of canonical variables
\begin{equation}
\begin{split}
&\hat {q}_{\alpha}(t) = \sqrt{\frac{\hbar}{2 m_{\alpha} \omega_{\alpha}}} \left[\hat{a}^{+}_{\alpha}(t) + \hat{a}_{\alpha}(t)\right]\\
&\hat {p}_{\alpha}(t) = i \sqrt{\frac{\hbar m_{\alpha} \omega_{\alpha}}{2}} \left[\hat{a}^{+}_{\alpha}(t) - \hat{a}_{\alpha}(t)\right]. 
\end{split}
\end{equation}
Then field function operators are
\begin{equation}
\hat{\vec{E}}(\vec{r},t) = \{\sum_{\alpha=1}^{\infty} \sqrt{\frac{\hbar \omega_{\alpha}}{V\epsilon_0}} \left[\hat{a}^{+}_{\alpha}(t) + \hat{a}_{\alpha}(t)\right] sin(k_{\alpha} z)\} \vec{e}_x,
\end{equation}

\begin{equation}
\hat{\vec{H}}(\vec{r},t) = i \{\sum_{\alpha=1}^{\infty} \sqrt{\frac{\hbar \omega_{\alpha}}{V\mu_0}} \left[\hat{a}^{+}_{\alpha}(t) - \hat{a}_{\alpha}(t)\right] cos(k_{\alpha} z)\} \vec{e}_y,
\end{equation} 
For the second partial solution, corresponding to Hamiltonian  $\mathcal{H}^{[2]}(t)$ we have
\begin{equation}
\begin{split}
&\left[\hat{p}{''}_{\alpha}(t) , \hat {q}{''}_{\beta}(t)\right] = i\hbar\delta_{{\alpha}\beta}\\
&\left[\hat {q}{''}_{\alpha}(t) , \hat {q}{''}_{\beta}(t)\right] = \left[\hat {p}{''}_{\alpha}(t) , \hat {p}{''}_{\beta}(t)\right] = 0,
\end{split}
\end{equation}
$\alpha, \beta \in N$.
The operators $\hat{a}{''}_{\alpha}(t)$, $\hat{a}{''}^{+}_{\alpha}(t)$ are introduced analogously
\begin{equation}
\begin{split}
&\hat{a}{''}_{\alpha}(t) = \frac{1}{ \sqrt{ 2 \hbar  m_{\alpha} \omega_{\alpha}}} \left[ m_{\alpha} \omega_{\alpha}\hat {q}{''}_{\alpha}(t) + i \hat {p}{''}_{\alpha}(t)\right]\\
&\hat{a}{''}^{+}_{\alpha}(t) = \frac{1}{ \sqrt{ 2 \hbar  m_{\alpha} \omega_{\alpha}}} \left[ m_{\alpha} \omega_{\alpha}\hat {q}{''}_{\alpha}(t) - i \hat {p}{''}_{\alpha}(t)\right]
\end{split}
\end{equation}
Relationships for canonical variables are
\begin{equation}
\begin{split}
&\hat {q}{''}_{\alpha}(t) = \sqrt{\frac{\hbar}{2 m_{\alpha} \omega_{\alpha}}} \left[\hat{a}{''}^{+}_{\alpha}(t) + \hat{a}{''}_{\alpha}(t)\right]\\
&\hat {p}{''}_{\alpha}(t) = i \sqrt{\frac{\hbar m_{\alpha} \omega_{\alpha}}{2}} \left[\hat{a}{''}^{+}_{\alpha}(t) - \hat{a}{''}_{\alpha}(t)\right] 
\end{split}
\end{equation}
 For the field function operators we obtain
\begin{equation}
\begin{split}
&\hat{\vec{E}}^{[2]}(\vec{r},t) = \\
&i \{\sum_{\alpha=1}^{\infty} \sqrt{\frac{\hbar \omega_{\alpha}}{V\epsilon_0}} \left[\hat{a}{''}^{+}_{\alpha}(t) - \hat{a}{''}_{\alpha}(t)\right] sin(k_{\alpha} z)\} \vec{e}_x,
\end{split}
\end{equation}
\begin{equation}
\begin{split}
&\hat{\vec{H}}^{[2]}(\vec{r},t) = \\
&\{\sum_{\alpha=1}^{\infty} \sqrt{\frac{\hbar \omega_{\alpha}}{V\mu_0}} (-1) \left[\hat{a}{''}^{+}_{\alpha}(t) + \hat{a}{''}_{\alpha}(t)\right] cos(k_{\alpha} z)\} \vec{e}_y.
\end{split}
\end{equation}
Let us designate the vector-functions of the first partial solution with index $[1]$. In accordance with definition of complex quantities we have
\begin{equation}
(\vec{E}^{[1]}(\vec{r},t), \vec{E}^{[2]}(\vec{r},t)) \rightarrow \vec{E}^{[1]}(\vec{r},t) +  i \vec{E}^{[2]}(\vec{r},t) = \vec{E}(\vec{r},t),
\end{equation}
\begin{equation}
(\vec{H}^{[2]}(\vec{r},t), \vec{H}^{[1]}(\vec{r},t)) \rightarrow \vec{H}^{[2]}(\vec{r},t) +  i \vec{H}^{[1]}(\vec{r},t) = \vec{H}(\vec{r},t).
\end{equation}
Consequently, correct field operators for quantized EM-field are
\begin{equation}
\begin{split}
&\hat{\vec{E}}(\vec{r},t) =  \{\sum_{\alpha=1}^{\infty} \sqrt{\frac{\hbar \omega_{\alpha}}{V\epsilon_0}} \{\left[\hat{a}^{+}_{\alpha}(t) + \hat{a}_{\alpha}(t)\right]\\
& + \left[\hat{a}{''}_{\alpha}(t) - \hat{a}{''}^{+}_{\alpha}(t)\right]\} sin(k_{\alpha} z)\} \vec{e}_x,
\end{split}
\end{equation}
and
\begin{equation}
\begin{split}
&\hat{\vec{H}}(\vec{r},t) =  \{\sum_{\alpha=1}^{\infty} \sqrt{\frac{\hbar \omega_{\alpha}}{V\mu_0}}\{ \left[\hat{a}^{+}_{\alpha}(t) - \hat{a}_{\alpha}(t)\right] \\
&- \left[\hat{a}{''}_{\alpha}(t) + \hat{a}{''}^{+}_{\alpha}(t)\right] \} cos(k_{\alpha} z) \} \vec{e}_y,
\end{split}
\end{equation}
\section{Quantization of EM-field in volume rectangular cavity with space coordinate dependent photon creation and annihilation operators}

Suppose EM-field also in volume rectangular cavity and that the field polarization is linear in z-direction. Then the vector of electrical component can be represented also in the form 
\begin{equation}
E_x(z,t)\vec{e}_x = \left[ \sum_{\alpha=1}^{\infty}A'_{\alpha}q_{\alpha}(z)\sin(\omega_{\alpha}t)\right] \vec{e}_x,
\end{equation}
where $q_{\alpha}(t)$ is  $\alpha$-th normal mode of the cavity, $\alpha \in N$, 
\begin{equation}
k_{\alpha} = \frac{\alpha\pi}{L},  A'_{\alpha}=\sqrt{\frac{2 \omega_{\alpha}^2}{T}}, \omega_{\alpha} = \frac{\alpha \pi c}{L},
\end{equation}
$L$ is cavity length along z-axis, $T$ is fixed time value. Using the equation
\begin{equation}
\epsilon_0\partial_t \vec{E}(z,t) = \left[\nabla\times\vec{H}(z,t)\right]
\end{equation}
we obtain the expression for magnetic field    
\begin{equation}
\begin{split}
& {H}_y(z,t)\vec{e}_y = \\
& \left\{ - \sum_{\alpha=1}^{\infty}\left[ A_{\alpha}\omega_{\alpha}\left( \int\limits_{0}^{z} q_{\alpha}(z') dz'\right) \cos(\omega_{\alpha}t) + H_{y0}(t)\right] \right\} \vec{e}_y,
\end{split}
\end{equation}
where $H_{y0} = \sum_{\alpha=1}^{\infty} f_{\alpha}(t)$, $\{f_{\alpha}(t)\}$, $\alpha \in N$, is the set of arbitrary  functions of the time.
We will use the partial solution, in which the function $H_{y0}(t)$ is identically zero. It is convenient to use new variable
\begin{equation}
\int\limits_{0}^{z} q_{\alpha}(z') dz' = q'_{\alpha}(z),
q_{\alpha}(z) = \frac{dq'_{\alpha}(z)}{dz}
\end{equation}
Then, analogously to Hamiltonian, we can consider the quantity
\begin{equation}
G(z) = \frac{1}{2}\int\limits_{0}^{t}\left[ \vec{E}^2(z,\tau) + \vec{H}^2(z,\tau)\right] d\tau,
\end{equation}
which represents itself the corresponding component of energy-impulse tensor (CGS-system is used).
Taking into account the expresions for $ E_x(z,t), {H}_y(z,t)$ we have
\begin{equation}
\begin{split}
&G(z) = \frac{1}{2}\int\limits_{0}^{t}\left\{\sum_{\alpha=1}^{\infty}{A'}_{\alpha}^{2}\left[\frac{dq_{\alpha}'(z)}{dz}\right]^{2} \sin^2\omega_{\alpha}\tau\right\}d\tau\\
&+ \int\limits_{0}^{t}\left\{\sum_{\alpha=1}^{\infty}\sum_{\beta=1}^{\infty}{A'}_{\alpha}{A'}_{\beta} \frac{dq_{\alpha}'(z)}{dz}\frac{dq_{\beta}'(z)}{dz}\sin\omega_{\alpha}\tau\sin\omega_{\beta}\tau\right\}d\tau \\
&+ \frac{1}{2}\int\limits_{0}^{t}\left\{  \sum_{\alpha=1}^{\infty}{A'}_{\alpha}^{2}\omega^{2}_{\alpha}\left[q_{\alpha}'(z)\right]^{2}\cos^2\omega_{\alpha}\tau\right\}d\tau
\\ &+ \int\limits_{0}^{t}\left\{ 
\sum_{\alpha=1}^{\infty}\sum_{\beta=1}^{\infty}{A'}_{\alpha}{A'}_{\beta} q_{\alpha}'(z)q_{\beta}'(z)\cos\omega_{\alpha}\tau\cos\omega_{\beta}\tau\right\}d\tau
\end{split}
\end{equation}
If we integrate, then we obtain
\begin{equation}
G(z) = \frac{1}{2}\sum_{\alpha=1}^{\infty}\left\{\omega^{2}_{\alpha}\left[\frac{dq_{\alpha}'(z)}{dz}\right]^{2} + \omega^{4}_{\alpha}\left[q_{\alpha}'(z)\right]^{2}\right\}
\end{equation}
Let us introduce the variables
\begin{equation}
\begin{split}
&q_{\alpha}''(z) = \omega_{\alpha} q_{\alpha}'(z),\\
&p_{\alpha}''(z) = \omega_{\alpha} \frac{dq_{\alpha}'(z)}{dz},
\end{split}
\end{equation}
then $G(z)$ in canonical form is
\begin{equation}
G(z) = \frac{1}{2}\sum_{\alpha=1}^{\infty}\left\{ \left[p_{\alpha}''(z)\right]^{2} + \omega^{2}_{\alpha}\left[q_{\alpha}^{''}(z)\right]^{2}\right\}. 
\end{equation}
Quantization is realized, if to take into account the relationships
\begin{equation}
\begin{split}
&\left[\hat{p}{''}_{\alpha}(z) , \hat {q}{''}_{\beta}(z)\right] = i\lambda_{0}\delta_{{\alpha}\beta}\\
&\left[\hat {q}{''}_{\alpha}(z) , \hat {q}{''}_{\beta}(z)\right] = \left[\hat {p}{''}_{\alpha}(z) , \hat {p}{''}_{\beta}(z)\right] = 0,
\end{split}
\end{equation}
where $\alpha,\beta \in N$.
The operators $\hat{a}{''}_{\alpha}(z)$, $\hat{a}{''}^{+}_{\alpha}(z)$ are introduced analogously to operators $\hat{a}{''}_{\alpha}(t)$, $\hat{a}{''}^{+}_{\alpha}(t)$ and they are
\begin{equation}
\begin{split}
&\hat{a}{''}_{\alpha}(z) = \frac{1}{ \sqrt{ 2  \lambda_{0}  \omega_{\alpha}}} \left[ \omega_{\alpha}\hat {q}{''}_{\alpha}(z) + i \hat {p}{''}_{\alpha}(z)\right]\\
&\hat{a}{''}^{+}_{\alpha}(z) = \frac{1}{ \sqrt{ 2  \lambda_{0} \omega_{\alpha}}} \left[ \omega_{\alpha}\hat {q}{''}_{\alpha}(z) - i \hat {p}{''}_{\alpha}(z)\right].
\end{split}
\end{equation}
Relationships for operators of canonical variables $\hat {q}{''}_{\alpha}(z)$ and $\hat {p}{''}_{\alpha}(z)$ are
\begin{equation}
\begin{split}
&\hat {q}{''}_{\alpha}(z) = \sqrt{\frac{\lambda_{0}}{2 \omega_{\alpha}}} \left[\hat{a}{''}^{+}_{\alpha}(z) + \hat{a}{''}_{\alpha}(z)\right]\\
&\hat {p}{''}_{\alpha}(z) = i \sqrt{\frac{\lambda_{0} \omega_{\alpha}}{2}} \left[\hat{a}{''}^{+}_{\alpha}(z) - \hat{a}{''}_{\alpha}(z)\right]. 
\end{split}
\end{equation}
It is easily to show, that the operator $\hat{G}(z)$ can be represented in the simple form
\begin{equation}
\hat{G}(z) = \sum_{\alpha=1}^{\infty}\lambda_{0} \omega_{\alpha}\left[\hat{a}{''}^{+}_{\alpha}(z)\hat{a}{''}_{\alpha}(z) + \frac{1}{2}\right],
\end{equation}
which determines physical meaning of the operators
$\hat{a}{''}^{+}_{\alpha}(z)$ and $\hat{a}{''}_{\alpha}(z)$.
The operators of vector-functions of EM-field can be represented in the form
\begin{equation}
\hat{\vec{E}}(\vec{r},t) = \left\{i\sum_{\alpha=1}^{\infty} A'_{\alpha} \sqrt{\frac{\lambda_{0}}{2 \omega_{\alpha}}} sin\omega_{\alpha}t \left[\hat{a}{''}^{+}_{\alpha}(z) - \hat{a}{''}_{\alpha}(z)\right] \right\} \vec{e}_x
\end{equation}
for $\hat{\vec{E}}(\vec{r},t)$ and
\begin{equation}
\hat{\vec{H}}(\vec{r},t) = \left\{\sum_{\alpha=1}^{\infty} A'_{\alpha} \sqrt{\frac{\lambda_{0}}{2 \omega_{\alpha}}} cos\omega_{\alpha}t \left[\hat{a}{''}_{\alpha}(z) + \hat{a}{''}^{+}_{\alpha}(z)\right] \right\}     \vec{e}_y
\end{equation}
for $\hat{\vec{H}}(\vec{r},t).$
\section{Local Quantization of EM-field}

Let us consider the field vector-functions in general form. So, if
$\vec{E}(\vec{r},t)$ is
\begin{equation}
\vec{E}(\vec{r},t) = \left\{\sum_{\alpha=1}^{\infty} {A''}_{\alpha} q_{\alpha}(t) q_{\alpha}(z)\right\}\vec{e}_x,
\end{equation}
then 
$\vec{H}(\vec{r},t)$ will be
\begin{equation}
\vec{H}(\vec{r},t) = \left\{-\sum_{\alpha=1}^{\infty} {A''}_{\alpha} \frac{dq_{\alpha}(t)}{dt}\left(\int\limits _{0}^{z} q_{\alpha}(z')dz'\right) \right\}\vec{e}_y.
\end{equation}
It represents the interest to get the expression for the density of the components of energy-impulse tensor $W(\vec{r},t)$, that is for the quantity
\begin{equation}
W(\vec{r},t) =  \frac{1}{2}\left[\epsilon_{0}\vec{E}^2(z,t) + \mu_{0} \vec{H}^2(z,t)\right]
\end{equation}
through the electric and magnetic fields
\begin{equation}
\begin{split}
&W(\vec{r},t) = \frac{1}{2}\left\{\sum_{\alpha=1}^{\infty}\frac{{A''}_{\alpha}^{2}}{\omega_{\alpha}^2}\left[{p''}_{\alpha}(z)\right]^{2} \left[q_{\alpha}(t)\right]^{2}\right\}\\
&+ \frac{1}{2}\left\{\sum_{\alpha=1}^{\infty}\sum_{\beta=1}^{\infty}2\frac{{A''}_{\alpha}{A''}_{\beta}}{\omega_{\alpha}\omega_{\beta}} {p''}_{\alpha}(z){p''}_{\beta}(z)q_{\alpha}(t)q_{\beta}(t)\right\} \\
&+ \frac{1}{2}\left\{\sum_{\alpha=1}^{\infty}\frac{{A''}_{\alpha}^{2}}{\omega_{\alpha}^2}\left[{p''}_{\alpha}(t)\right]^{2} \left[q_{\alpha}(z)\right]^{2}\right\}
\\ &+\frac{1}{2}\left\{\sum_{\alpha=1}^{\infty}\sum_{\beta=1}^{\infty}2\frac{{A''}_{\alpha}{A''}_{\beta}}{\omega_{\alpha}\omega_{\beta}} {p''}_{\alpha}(t){p''}_{\beta}(t)q_{\alpha}(z)q_{\beta}(z)\right\} 
\end{split}
\end{equation}
We consider the case of diagonalized quadratic form, in particular, when in foregoing expression two nondiagonal terms are zero. Then, introducing the operators
\begin{equation}
\begin{split}
&\hat{a}_{\alpha}(z,t) = \\
&\frac{c \omega_{\alpha}}{\sqrt{2 (\lambda'_{0} + \hbar)  \omega_{\alpha}}} \left[ \omega_{\alpha} c \hat {q''}_{\alpha}(z) + i c \hat {p''}_{\alpha}(z)\right] \left[ \omega_{\alpha}\hat {q}_{\alpha}(t) + i \hat {p}_{\alpha}(t)\right]\\
&\hat{a}^{+}_{\alpha}(z,t) = \\
&\frac{c \omega_{\alpha}}{\sqrt{2 (\lambda'_{0} + \hbar)  \omega_{\alpha}}} \left[ \omega_{\alpha} c \hat{q''}_{\alpha}(z) - i c \hat {p''}_{\alpha}(z)\right] \left[ \omega_{\alpha}\hat {q}_{\alpha}(t) - i \hat {p}_{\alpha}(t) \right],
\end{split}
\end{equation}
where $\lambda'_{0} = \frac{\lambda_{0}}{c}$. 
It can be shown, that operators $\hat{a}_{\alpha}(z,t)$ and $\hat{a}^{+}_{\alpha}(z,t)$ satisfy to commutation conditions
\begin{equation}
\left[\hat{a}_{\alpha}(z,t), \hat{a}^{+}_{\alpha}(z,t)\right] = i(\lambda'_{0} + \hbar) \mathcal{H}_{\alpha}(z,t), \alpha \in N.
\end{equation}
Here $\mathcal{H}_{\alpha}(z,t)$ is 4-density of the Hamiltonian of the $\alpha-th$ mode. In particular, if ${q''}_{\alpha}(z), \alpha \in N$, are harmonical functions, that is in the case, corresponding to the most of the practical applications (taking into account the possibility of expansion in Fourier 
series) the commutation relationships are
\begin{equation}
\left[\hat{a}_{\alpha}(z,t), \hat{a}^{+}_{\alpha}(z,t)\right] = i(\lambda'_{0} + \hbar), \alpha \in N.
\end{equation}
So we obtain the rules for local quantization of EM-field in Minkowski space. Given consideration shows, that simultaneously with time of creation or annihilation of the photons it becomes to be possible to determine the place of photon creation or annihilation. 
\section{Discussion and Conclusions}

\end{document}